\documentclass[%
reprint,
 amsmath,amssymb,
 aps,
]{revtex4-1}
\usepackage{epsfig,epsf,rotating,wrapfig,tabularx}
\usepackage{dcolumn}
\usepackage{bm}
\usepackage{graphicx}

\begin{document}

\preprint{APS/123-QED}

\title{Comparative Analysis of Pion, Kaon and Proton Spectra Produced at PHENIX}

\author{\firstname{A.~A.}~\surname{Bylinkin}}
 \email{alexander.bylinkin@desy.de}
\affiliation{%
 Institute for Theoretical and Experimental
Physics, ITEP, Moscow, Russia
}%
\author{\firstname{A.~A.}~\surname{Rostovtsev}}
 \email{rostov@itep.ru}
\affiliation{%
 Institute for Theoretical and Experimental
Physics, ITEP, Moscow, Russia
}%


\begin{abstract}
The shapes of invariant differential cross section for identified $\pi^{\pm}$,$K^{\pm}$, $p$ and $\overline{p}$ production as function of transverse momentum measured in $pp$ collisions by the PHENIX detector are analyzed. Simultaneous fit of these data to the sum of exponential and power-law terms show significant difference in the exponential term contributions. This effect qualitatively explains the observed shape of the experimental $K/\pi$ and $p/\pi$ yield ratios measured as function of transverse momentum of produced hadrons. A picture with two types of mechanisms for hadron production is given.
\end{abstract}

\pacs{Valid PACS appear here}
\maketitle


There is a large volume of experimental data available on charged particle and identified hadrons production in high energy particle collisions collected starting from the very first experiments performed on the Intersecting Storage Rings (ISR) at CERN and till contemporary high statistics measurements carried out at RHIC and LHC. In most available publications the measured experimental spectra of produced hadrons become a subject for a phenomenological description or for a comparison with selected models in each experiment separately. However, a comparative simultaneous analysis of the whole available data volume could provide a new powerful lever arm to disclose a common underlying dynamics in work in the hadron production in high energy particle collisions. Such systematic analysis have been described in two our recent papers~\cite{OUR1,OUR2}.

In~\cite{OUR1} it was demonstrated that spectra of charged hadrons produced in collisions of baryons require an existence of a sizable fraction of an exponential (Boltzmann-like) statistical ensemble of charged hadrons on the top of a power-law (pQCD inspired spectrum shape) functional term. 
According to~\cite{OUR1} the overall generic charged hadron spectrum as function of the produced hadron transverse momentum ($P_T$) is given by a sum of the exponential and power-law terms
\begin{equation}
\label{eq:exppl}
\frac{d\sigma}{P_T d P_T} = A_e\exp {(-E_{Tkin}/T_e)} +
\frac{A}{(1+\frac{P_T^2}{T^{2}\cdot n})^n},
\end{equation}
where  $E_{Tkin} = \sqrt{P_T^2 + M^2} - M$
with M equal to the produced hadron mass. $A_e, A, T_e, T, n$ are the free parameters to be determined by fit to the data.

\begin{figure}[h]
\includegraphics[width =8cm]{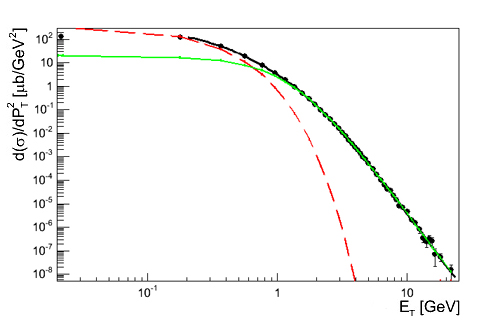}
\caption{\label{fig:01} Charged particle spectrum~\cite{UA1} fitted to the function~(\ref{eq:exppl}): the red (dashed) curve shows the exponential term and the green (solid) one stands for the power law term.}
\end{figure}

A typical charged particle spectrum as function of transverse energy, fitted to the function~(\ref{eq:exppl}) is shown in Fig~\ref{fig:01}. In addition the contributions off the exponential and power law terms to the spectrum are shown in Fig~\ref{fig:01} separately. It is observed, that the exponential Boltzman-like term dominates the charge particle spectrum at low $P_T$ values.
Since the absolute majority of produced charged hadrons are charged $\pi$ it was concluded that the observed charged hadron spectrum shape is also characteristic for charged $\pi$. This conclusion was supported by the studies of the identified charged $\pi$ production spectra~\cite{OUR1}.

Though the data on the identified hadron production are sparse and available in the very limited range of the hadron transverse momentum only it is still interesting to analyse the spectra of $K$ and $p$ in a way proposed in~\cite{OUR1}. This analysis was performed in~\cite{OUR2}. It turned out that, contrary to the pions, the spectra of $K$ and $p$ leave a relatively little room for the exponential term contribution and within the experimental errors could be described by the power law term only. 

In the present paper an attempt was made to make a simultaneous comparative analysis of the $\pi$,$K$ and $p$ spectra produced at the same collision energy and under the same experimental conditions. For this comparison the data from the detector PHENIX collected in $pp$ collision run at $\sqrt(s)=200 GeV$ were used~\cite{Adare:2011vy}. 

\begin{figure*}[!]
\includegraphics[width = 18cm]{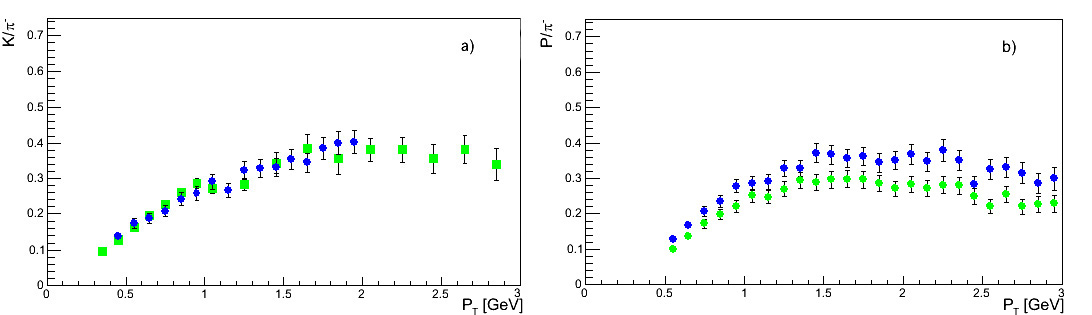}
\caption{\label{fig:02}  Ratios of the measured differential cross sections $K/\pi$ (a) and $p/\pi$ (b) measured at PHENIX~\cite{Adare:2011vy}. For (a): $K^{\pm}$ - blue points $K^{0}_{s}$ - green points. For (b): $p/\pi^{-}$ - blue points,  $\overline{p}/\pi^{-}$green points.}
\end{figure*}

\begin{figure*}[!]
\includegraphics[width = 18cm]{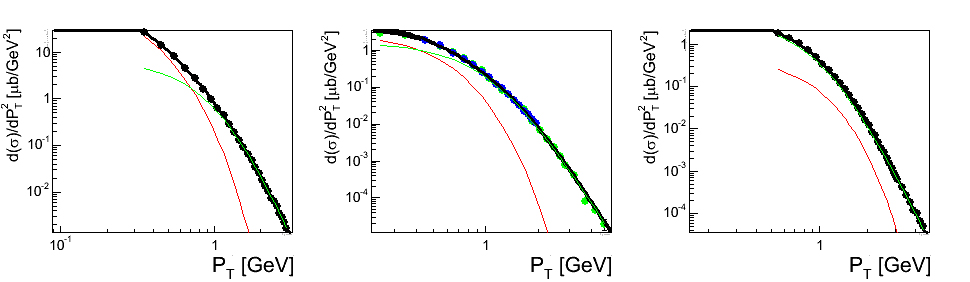}
\caption{\label{fig:03} $\pi$, $K$ and $p$ spectra~\cite{Adare:2011vy} fitted simultaneously: the red line shows the exponential term and the green one stands for the power law. The exponential term dominates for $\pi$ spectra only.}
\end{figure*}

As the first step one could note a peculiar behavior of the ratios of  differential cross sections $K/\pi$ and $p/\pi$ measured in the experiment. These ratios plotted as function of $P_T$ of produced hadrons are shown in Fig~\ref{fig:02}a, b. In both cases the ratio reaches a plateau above $P_T \approx 2 GeV$ and drops down for low $P_T$ values. Within the framework of the proposed approach based on the formula~(\ref{eq:exppl}) the  observation of this plateau suggests that parameter $n$ of the power law term in~(\ref{eq:exppl}) is likely to be similar both for $K$, $p$ and $\pi$.  In the QCD model this is correct at high  $P_T$ values since the produced hadron distributions are largely driven by the gluon momentum distribution in the colliding particles. 
If so, the identified $\pi$,$K$ and $p$ spectra are fitted simultaneously to the function~(\ref{eq:exppl}) with a constraint to have the same value for $n$ parameter in the power law term for all three spectra together. As the measured charged kaon spectrum is restricted to the low-$P_T$ values it was extended using the available $K^{0}_{s}$ data. The result of this fit procedure is shown in Fig~\ref{fig:03} and table~\ref{tab:01}. It is important to note that the values for the parameter $T$ obtained from such fit procedure turn out to be practically the same for all types of produced hadrons. Such surprising result was not obvious a priori. As it was expected the Boltzmann-like term for $\pi$ production dominates, while it gives much lower contribution in $K$ spectrum and it is practically close to zero for the $p$ spectrum. 
The relative amount of the exponential term contribution to the hadron spectra estimated from the simultaneous fit of $\pi$,$K$ and $p$ differential cross sections to the function~(\ref{eq:exppl}) are given in Table~1. It is important to note that the observed significant difference in the exponential contributions to the hadron spectra implies the difference in the hadron production mechanisms rather than an artifact of the fit procedure.    
\begin{figure*}[!]
\includegraphics[width = 18cm]{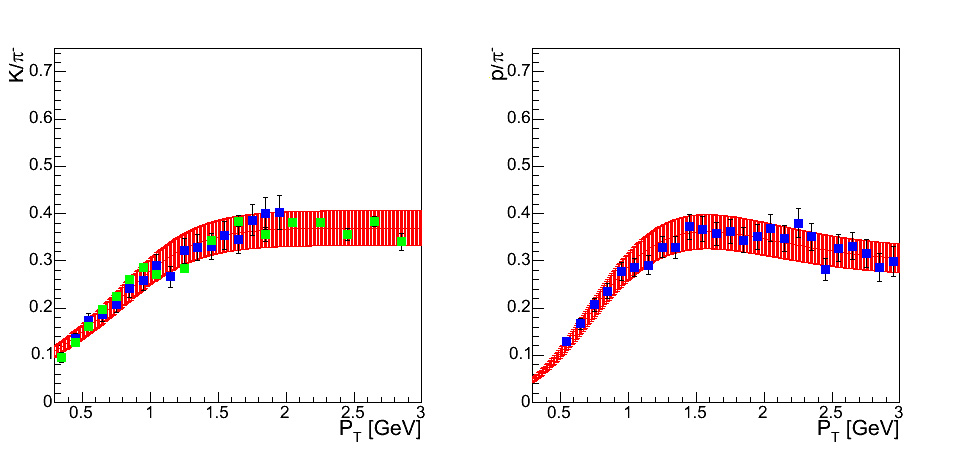}
\caption{\label{fig:04}$K/\pi$ and $p/\pi$ ratios for experimental data~\cite{Adare:2011vy} (points) shown over the corresponding ratios obtained from the results of the simultaneous fit (line). The error band is calculated using the parameter uncertainties obtained from the fit.}
\end{figure*}

\begin{table}
\begin{tabular}{|c|c|c|c|}
\hline
Type of  & Exponential & Power law& Exponential\\
produced hadron & contribution [\%] &Terms&Terms\\\hline
$\pi$ & $75\pm2$& - & - \\\hline
$K$ &  $28\pm7$&$0.29\pm0.03$ &$0.04\pm0.003$\\\hline
$p$ &$18\pm8$ &$0.25\pm0.02$ & $0.02\pm0.001$\\
\hline
\end{tabular}
\caption{Contribution of the exponential (Boltzmann-like) term to the charge particle spectra and $K/\pi$ and $p/\pi$ values calculated separately for power-law and exponential terms of~(\ref{eq:exppl}) obtained from the simultaneous fit of $\pi$,$K$ and $p$~\cite{Adare:2011vy}.}
\label{tab:01}
\end{table}

Once arrived to this conclusion, one could explain the peculiar behavior of the $K/\pi$ and $p/\pi$ ratios (Fig~\ref{fig:02}) as function of $P_T$ of produced particles. Figure~\ref{fig:04} shows that the ratios of differential cross sections $K/\pi$ and $p/\pi$ measured in the experiment and those obtained from the fit are in a good agreement with each other. The existence of large fraction of low-$P_T$ $\pi$ represented by the Boltzmann exponential statistical distribution suppresses these ratios at low  values of $P_T$. This situation is somewhat similar to that found in~\cite{OUR1} for $\pi$ production in $pp$ and heavy ion collisions taken at the same collision energy per nucleon. The ratio of the correspondent differential cross sections as function of $P_T$ significantly differs from unit and is traditionally interpreted as a signature of the nuclear absorption effect. It has been shown in~\cite{OUR1} that the shape of this ratio is formally defined by a difference of partial contribution of the exponential term to the sum~(\ref{eq:exppl}) for minimum bias $Au-Au$ and $pp$ interactions. 

Within the framework of the proposed approach the hadron production mechanism could be represented in a way of a sum of two different contributions:

1)	release of quasi-thermalized hadrons (mostly $\pi$, and much less probable $K$ and $p$), described by the exponential Boltzmann spectrum  and

2)	pQCD like production of hadrons described by the power law statistical distribution. The parameters of this distribution do not depend on the type of hadron produced in $pp$ collisions. 

It is worthwhile to note that as was found in~\cite{OUR1} a sizable contribution of the thermalized hadrons shows up in the baryonic collisions only. Whilst the interactions with the high energy photons involved as colliding particles do not require an extra exponential statistical distribution to describe the produced hadronic spectra.

In conclusion, it is found that the $\pi$ production in $pp$ collisions is dominated by a release of quasi-thermalized particles, while the spectra of heavier $K$, $p$ and $\overline{p}$ are dominated by pQCD like production mechanism, leaving a relatively small room for thermalized particle production. For $\pi$, $K$ and $p$ production the parameters of the power law contribution take practically the same (within the errors) values which depend on the global conditions like a collision energy, or a type of colliding particles rather than a type of produced hadron. 


\end{document}